# Further Thoughts on Abnormal Chromatin Configuration and Oncogenesis

Gao-De Li

Chinese Acupuncture Clinic, Liverpool, UK
Email: gaode_li@yahoo.co.uk

## Abstract

More than 30 years ago, we published a paper entitled as "Abnormal Chromatin Configuration and Oncogenesis", which proposed the first hypothesis that links oncogenesis to abnormal three-dimensional (3D) genome structure. Recently, many studies have demonstrated that the 3D genome structure plays a major role in oncogenesis, which strongly supports our hypothesis. In this paper, further thoughts about our hypothesis are presented.

## Subject Areas

Cell Biology, Developmental Biology, Genomics, Molecular Biology

## Keywords

Chromatin Configuration, Oncogenesis, 3D Genome, Cancer, Cell Differentiation, Cell Cycle, Gene Regulation, Gene Activity

## 1. Introduction

In 1986, we published a paper entitled as "Abnormal Chromatin Configuration and Oncogenesis" in a Chinese Journal named as Medicine and Philosophy [1], which proposed the first hypothesis that links oncogenesis to abnormal three-dimensional (3D) genome structure. At the time when we published this hypothesis, exploration of the 3D chromatin configuration and gene regulation was impractical because no suitable techniques and genome DNA sequence database available. In 2002, chromosome conformation capture technique was invented [2], which opens a new field of the 3D genome structure research. Since then more advanced techniques have been developed [3]. Thanks to these powerful techniques, a lot of progress has been made in recent 10 years. From 2015 onwards, more and more studies have demonstrated that chromatin configuration is involved in gene regulation, cell differentiation, and plays a major role in oncoge-nesis [4]-[9]. We are very pleased to see that after nearly 3 decades, our hypothesis eventually gained strong support from experimental evidence. In this paper, further thoughts about our hypothesis are presented.

## 2. Key Points of our Hypothesis

Since our hypothesis was published in Chinese [1], which might be difficult for non-Chinese speakers to read. Therefore, an English summary of the key points of our hypothesis is presented as follows:

The spatial or 3-dimensional structure of entire chromatin fibers inside the nucleus of a eukaryotic cell could be defined as chromatin configuration which is associated with gene activity or determines the gene activity pattern. Based on this assumption, cell differentiation could be defined as a process in which different cells are endowed with different chromatin configurations so that gene activity patterns in different types of cells are different. Chromatin configuration is a highly ordered structure and constantly changes during cell cycle progression. Carcinogens can cause chromatin-structure change, and long-time exposure to carcinogens will result in irreversible abnormalities of chromatin configuration which brings about aberrant gene activity or cancer associated gene activity. Based on this hypothesis, a cell with gene mutations but without fundamental change in its chromatin configuration will have no chance to become a cancer cell. Furthermore, if normal chromatin configuration could be restored, cancer cells are likely to be turned back to normal.

## 3. Detailed Explanation of our Hypothesis

The nucleus of a living eukaryotic cell could be imagined to be a spherical space in which chromatin fibers from all chromosomes (human somatic cell has 46 chromosomes) interact each other and form a complicated 3Dgenome structure which is defined as chromatin configuration. We think that chromatin configuration might determine gene activity patterns and involved in gene regulation. Furthermore, we speculate that cell differentiation is a process in which different types of cells are endowed with different chromatin configurations so that they have different gene activity patterns that determine cell's phonotypes. The "blueprint" of chromatin configuration is stored in the genome.

Chromatin configuration is dynamically changing during cell cycle progression, which might have two main functions: one is to constantly regulate gene expression that is required by cell cycle progression, and the other is to response to any environmental change. Gene activities and chromatin configuration might be able to regulate each other, leading to a chain reaction-like regulation pattern during cell cycle progression.

Different types of cells have different chromatin configurations at different points of cell cycle, which is associated with their different gene activity patterns. During cell cycle progression, dynamic change of chromatin configuration starts from the point when chromosomes in new daughter cells just begin to uncoil to G.-D. Libecome chromatin fibers, and finishes at the point when chromatin is repackaged into chromosomes in the mother cell. Chromosome is a 3D structure composed of highly coiled and condensed chromatin fibers, which makes chromosome unsuitable for regulating gene activity. Besides, appearance of chromosomes indicates that chromatin configuration has been disassembled. The purpose of chromatin configuration regulating gene activity during cell cycle progress perhaps is to produce chromosomes which function like gift bags filled with chromatin fibers and are distributed into new daughter cells during cell division. In a word, chromosome is not involved in regulation of gene expression during cell cycle progression.

The foundation of chromatin configuration is DNA and proteins such as histones etc. Carcinogens (physical, chemical and biological) can directly or indirectly damage DNA and chromatin configuration associated proteins, leading to regional chromatin configuration change due to minor damage. At early stage, this change is reversible, but if exposure to carcinogens persists for a long term, accumulated damage will cause chromatin-configuration fundamental change that is fatal and hardly irreversible. Under this condition, majority of the cells with this abnormal chromatin configuration died, only one or few cells (clones) survived but with cancer phenotype, such as uncontrolled growth, evasion of apoptosis, tissue invasion and metastasis etc. [10]. This is how chromatin configuration is involved in oncogenesis.

Cancer cell's abnormal chromatin configuration could be defined as cancer associated chromatin configuration (CACC), the "blueprint" of which is installed in genome as a life-saving equipment during evolution. CACC determines cancer cell's gene activity pattern that determines cancer cell's phenotype. Compared to normal cell's chromatin configuration, CACC is unstable and easily affected by environmental change, which gives cancer cell an ability to fine tune its gene activities so that it can survive harmful environment. This is the reason why radiotherapy and chemotherapy usually can't completely kill all cancer cells in human body, leading to cancer recurrence. In addition, due to unstable CACC, cancer cells might constantly undergo clonal selections, resulting in heterogeneity in cancer phenotypes [11]. Some selected clones might be responsible for metastasis which causes most cancer deaths.

If cancer is caused by CACC, some new strategies could be developed to treat cancer, for example, using less-toxic drugs to turn CACC to normal so that cancer cells could become normal cells. If the strategy does not work, partial alteration of CACC by drugs might be possible, which could enable cancer cells become less harmful, for example, no metastasis occurs.

In conclusion, according to our hypothesis, only cells with CACC will become cancer cells. Any cells without CACC will not become cancer cells even if they have abnormal chromosomes, or virus infection, or mutations. Therefore, identification of fundamental structure of CACC in cancer cells is of paramount importance in winning the battle against cancer.

DOI: 10.4236/oalib.1105185        3        Open Access Library Journal

## 4. Implications of Our Hypothesis

Our hypothesis may have significant implications as it could be used to explain the mechanisms of both oncogenesis and cell differentiation, which may help to develop new strategies to treat cancer. It may also have implications in understanding of age-related disorders as chromatin configuration associated gene regulation may become inefficient in the cells of old people.

## 5. Conclusion

The paper we published in 1986 is of great importance in three aspects: first, it is the first published paper exploring the relationship between chromatin configuration (3D genome structure) and gene activity; second, it proposed the first hypothesis that links oncogenesis to abnormal 3D genome structure; third, it presented a speculation that cell differentiation is associated with 3D genome structure. We are very pleased to see that after nearly 30 years, all these viewpoints eventually gained strong support from research findings published in recent years. We believe that further investigation into the relationship between chromatin configuration and gene activity will contribute significantly to the development of life sciences.

## Conflicts of Interest

The author declares that there is no conflict of interest regarding the publication of this paper.